\documentclass[a4paper,11pt]{article}
\pdfoutput=1 

\usepackage{jinstpub} 

\usepackage{soul}

\usepackage{hyperref}

\usepackage{lineno}
\usepackage{multirow}
\usepackage{caption}
\usepackage{subcaption}
\usepackage{eqnarray,amsmath}

\usepackage[colorinlistoftodos]{todonotes}
\usepackage{soul,xcolor}

\setstcolor{red}

\title{Demonstrating a single-block 3D-segmented plastic-scintillator detector}

\author[a]{A.~Boyarintsev,}
\author[b]{A.~De Roeck,}
\author[b]{S.~Dolan,}
\author[c]{A.~Gendotti,}
\author[a]{B.~Grynyov,}
\author[b]{U.~Kose,}
\author[a]{S.~Kovalchuk,}
\author[a]{T.Nepokupnaya,}
\author[c]{A.~Rubbia,}
\author[c,1]{D.~Sgalaberna,}
\author[a]{T.~Sibilieva,}
\author[c]{and X.~Y.~Zhao}

\affiliation[a]{Institute for Scintillation Materials NAS of Ukraine (ISMA), Kharkiv 61072, Ukraine}
\affiliation[b]{European Organization for Nuclear Research (CERN), 1211 Geneva 23, Switzerland}
\affiliation[c]{ETH Zurich, Institute for Particle Physics and Astrophysics, CH-8093 Zurich, Switzerland}

\emailAdd{davide.sgalaberna@cern.ch}

\note{Corresponding author}

\abstract{
Three-dimensional finely grained plastic scintillator detectors bring many advantages in particle detectors, allowing a massive active target which enables a high-precision tracking of interaction products, excellent calorimetry and a sub-nanosecond time resolution. Whilst such detectors can be scaled up to several-tonnes, as required by future neutrino experiments, a relatively long production time, where each single plastic-scintillator element is independently manufactured and machined, together with potential challenges in the assembly, complicates their realisation. In this manuscript we propose a novel design for 3D granular scintillator detectors where O($1~\text{cm}^3$) cubes are efficiently glued in a single block of scintillator after being produced via cast polymerization, which can enable rapid and cost-efficient detector construction. This work could become particularly relevant for the detectors of the next-generation long-baseline neutrino-oscillation experiments, such as DUNE, Hyper-Kamiokande and ESSnuB.
}


\begin{document}
\maketitle
\flushbottom

\section{Introduction}
\label{sec:intro}


Organic scintillator detectors are widely used in particle physics experiments.
Neutrino detectors particularly benefit from the technology's high performance and scalability to large masses, thanks to the very-low production costs. Examples of their application can be found in the MINERvA~\cite{minerva}, MINOS~\cite{minos},
NOvA~\cite{NOvA:2019cyt} and T2K~\cite{t2k,t2k-fgd}  experiments. 
Other examples and applications can be found in sampling calorimeters
~\cite{calice,t2k-ecal}, 
fast neutron detection 
as well as in
muon~\cite{muon-tomography-pyramid} and positron-emission tomography~\cite{j-pet-1}.

Many organic scintillator detectors are polystyrene-based. In these detectors, an interacting charged particle excites the polymer matrix molecules and this energy is transferred to the activator, for example p-terphenyl (PTP), via a resonant dipole-dipole interaction (the Foerster mechanism~\cite{https://doi.org/10.1002/andp.19484370105}) which is inversely proportional to the 6-th power of the intramolecule distance. The interaction strongly couples the polymer base and the activator, such that a higher density of activator sharply increases the light yield of the plastic scintillator and reduces the light emission delay. 
A second dopant, for instance p-phenilene-bis(5-pheniloxazole) (POPOP), shifts the light wavelength to a region were the material is maximally transparent.
This light is then ultimately detected by photosensors, which count the number of scintillation photons produced by the interacting particle to permit a measurement of both the charged particle energy loss and the interaction time.
Particle tracking can then be enabled by, for example, optically-isolating certain volumes of the scintillator with optical reflectors or diffusers.

In recent years, neutrino experiments have begun to deploy 3D-granular plastic scintillator detectors with the goal of providing high-performance reconstruction of neutrino interactions.
Examples can be found in reactor~\cite{solid} and long-baseline neutrino oscillation (LBL) experiments~\cite{T2K:2019bbb}.
The ``SuperFGD''~\cite{superfgd}, employed by T2K's near detector upgrade, deploys three readout views to provide an isotropic three-dimensional reconstruction of neutrino interactions. 
The fine granularity additionally allows detection of low-energy protons and pions. 
In order to enhance the particle energy loss resolution, the scintillator 3D segmentation is provided by an optical reflector so that the scintillation light is trapped within a single scintillator volume and captured by wavelength-shifting (WLS) fibers. 
It should be noted that a scintillator with 3D optical segmentation requires a high transparency in order to avoid a strong correlation between the number of scintillation photons captured by the WLS fibers and the particle position and path inside the active volume, as such a correlation would distort attempts at particle identification.

A useful feature of organic scintillator for accelerator-based neutrino experiments is that the low mass carbon and hydrogen nuclei allow a relatively easy detection of the fast neutrons which are often produced in neutrino interactions. Neutrons with energies in the few to hundreds of MeV range tend to be able to either break up a carbon nucleus or eject the proton from hydrogen, leaving a clear experimental signature of the interaction~\cite{MINERvA:2019wqe}. If the time of both the neutrino interaction and the neutron scatter can be measured, the neutron energy can be determined~\cite{Munteanu:2019llq}. It is also possible to add nuclei such as lithium, boron or gadolinium to the scintillator such that thermalised neutrons can be efficiently detected~\cite{solid}.


Overall the features of organic scintillator detectors allow the necessary tracking and calorimetry performance to make them invaluable tools in the effort to better characterise neutrino interactions \cite{Dolan:2021hbw}. 
This in turn allows experiments to confront some of the most important and challenging systematic uncertainties faced by current and future accelerator-based neutrino oscillation experiments in their searches for leptonic CP violation and analyses of the neutrino mass hierarchy~\cite{Alvarez-Ruso:2017oui}.

Whilst the two-tonne SuperFGD represents a significant step forward in the deployment of organic scintillator technology for neutrino experiments, it is likely that more massive detectors will be required of future LBL experiments, DUNE~\cite{Abi:2020qib}, Hyper-Kamiokande~\cite{Hyper-Kamiokande:2018ofw} and ESSnuB~\cite{Wildner:2015yaa}, to achieve their physics goals, primarily the precise characterisation of Charge-Parity symmetry violation in neutrino oscillations. For instance, DUNE has in its Near Detector reference design a neutrino target analogous to SuperFGD but approximately five times more massive~\cite{DUNE:2021tad}. Hyper-Kamiokande will inherit the SuperFGD detector from the T2K experiment but may consider a further upgrade to increase the target mass. 

Although production and assembly methods of 
massive 3D granular scintillator
detectors have been developed and validated \cite{sfgd-testbeam-cern}, the scaling of such techniques to much larger volumes brings significant challenges. 
For instance,
the 
few-million optically-isolated scintillator cubes must be individually produced,
treated 
to produce a thin white reflector layer, 
drilled on three orthogonal faces to make holes for the WLS fibers, and individually placed inside a mechanical box. 
Such a construction procedure is time consuming both in the production and assembly steps.
Moreover, compiling a matrix of many millions of single cubes may cause stack up tolerance issues, that must be taken into account in the detector design. 

The ideal configuration would consist of a three-dimensional matrix of many optically isolated volumes all attached together to form a single block of scintillator. Such design would also be mechanically stronger than some million independent cubes.
This could potentially be achieved with additive manufacturing, 
where a matrix of plastic scintillator cubes is 3D printed \cite{Berns:2020ehg}.
However, more R\&D is necessary in order to make 3D printing of plastic scintillator sufficiently performant to build a large particle detector.

In this article we offer an alternative method to producing a large number of optically isolated elements within a single block of scintillator. The proposed design works from a standard scintillator production technique such as cast polymerization, altered to produce large layers of glued and optically-isolated scintillator cubes, setting the basis for the construction of future massive 3D fine-granularity plastic scintillator detectors. After detailing the new design in Sec.~\ref{sec:design}, we report the performance of a plastic scintillator prototype detector in Sec.~\ref{sec:measurements} and finally discuss the results as well as expected near-future developments and improvements in Sec.~\ref{sec:discussions}.

\section{Design and production of the 3D-segmented scintillator block}
\label{sec:design}

\begin{figure}
\centering
\raisebox{-0.5\height}{\includegraphics[width=5.07cm]{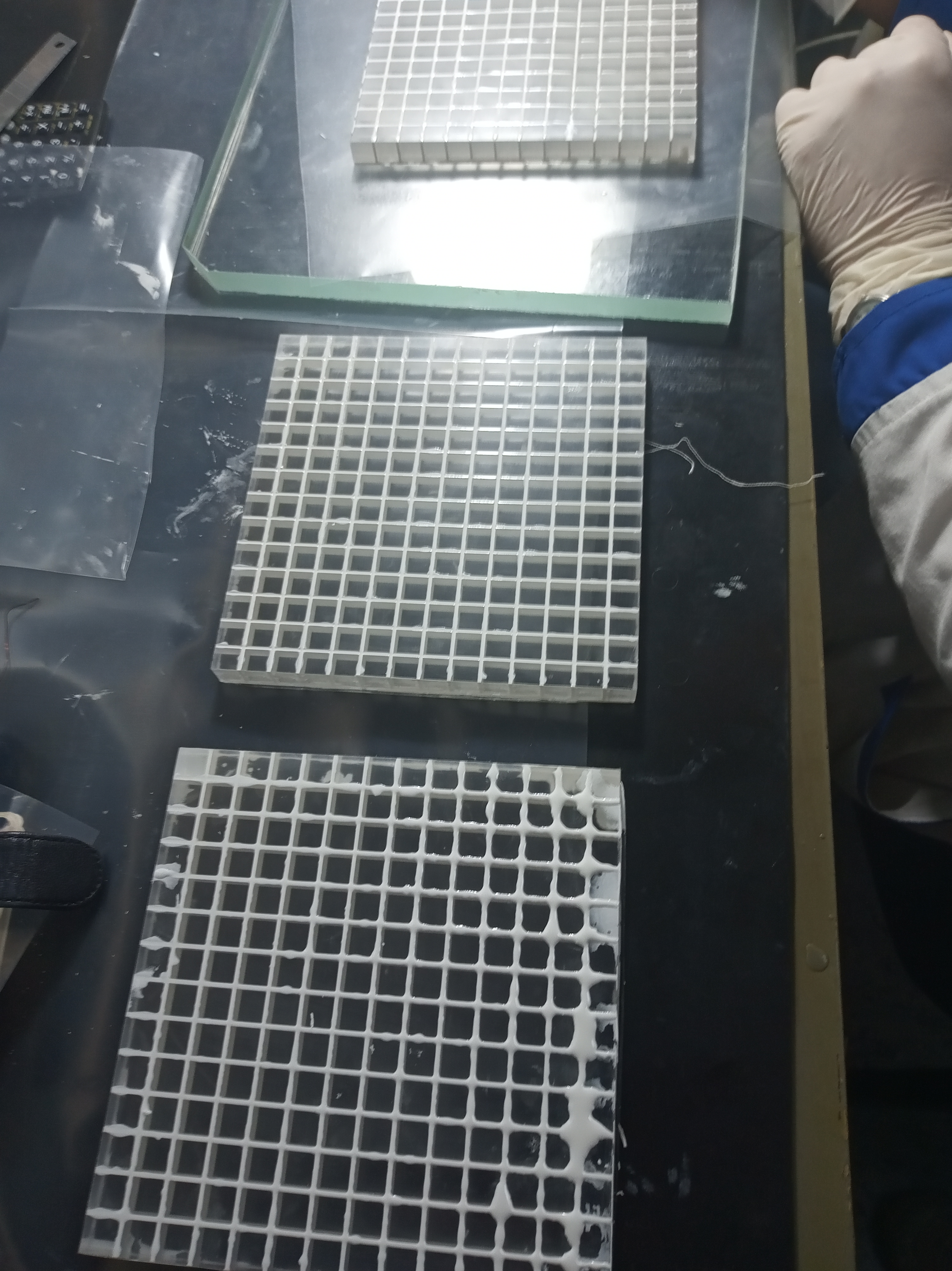}}
\raisebox{-0.5\height}{\includegraphics[width=9cm]{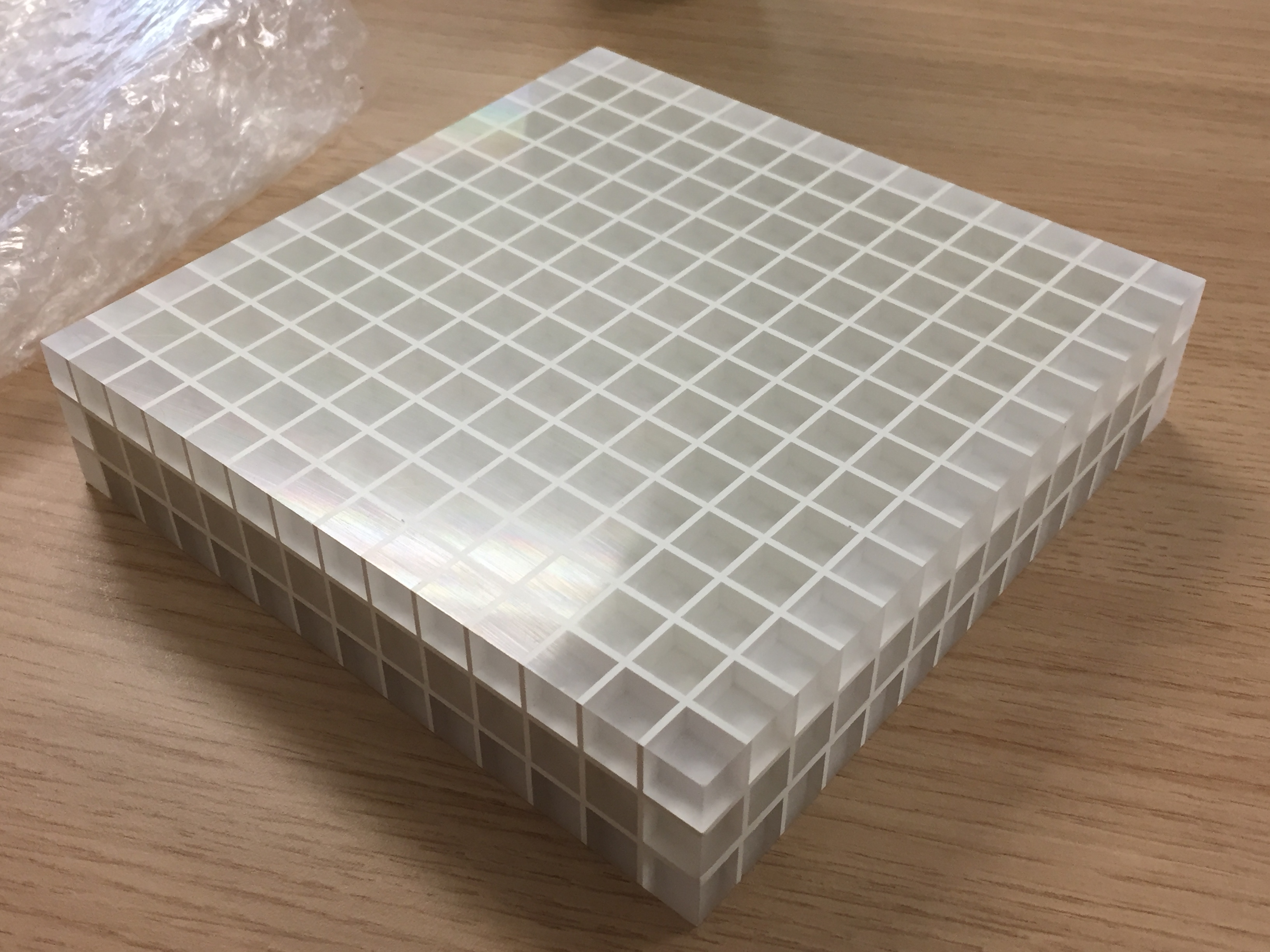}}
\caption{\label{fig:super-block} Left: gluing process of several scintillator cubes into different layers.
Right: manufactured scintillator volume made of $14 \times 14 \times 3$ optically-isolated cubes (588 in total) glued into a single solid block.
}
\end{figure}

The plastic scintillator considered in this work is UPS 923A and consists of polystyrene doped with 2\% by weight of PTP and 0.05\% by weight of POPOP.
A block of scintillator is produced with the cast polymerization technique~\cite{cast-polymerization}.
A liquid monomer with dissolved dopants is poured into a mold and heated. After cooling, a rigid solid plastic is obtained.
%
%
The produced scintillator layer is then processed with a CNC machine, using the array production technology: 1~mm gaps are made to define the matrix of cubes while all the elements are taking in position. Subsequently the gaps are filled with white-reflective epoxy modified resin.
The result is a matrix of optically-isolated cubes glued together.
%
%
The epoxy resin provides both the mechanical strength and rigidity to the 3D matrix as well as the optical isolation.
In Fig.~\ref{fig:super-block}, the 3D matrix is shown both during the gluing of the cubic volumes and after the completion of the manufacturing. 
It consists of a rigid block of 588 optically-isolated cubes ($14 \times 14 \times 3$), where each cube is made with an edge of 10~mm.
Eventually, the outermost surface can be polished and covered with $\text{TiO}_2$ paint.
The production and manufacturing of the scintillator matrices are carried out at the ``Institute for Scintillation Materials of the National Academy of Science of Ukraine'' (ISMA).

The procedure described above provides, for the first time, the unique opportunity to produce large matrices of finely-segmented scintillator volumes (up to several thousands) joined together in to a single rigid block. 
However, drilling several-centimeter long holes with O(1~mm) diameter is technically very hard to achieve.
We therefore adopt a design that consists of a stack of several individual layers of glued cubes, such that it is not necessary to drill long holes.
The three orthogonal WLS fibers can be inserted through two horizontal square grooves ($1.5 \times 1.6~\text{mm}^2$) and a vertical hole (1.5~mm diameter).
Tolerances on the hole and groove position, size of the cubes and the reflector thickness are between 0.1-0.2~mm.
A white tyvek sheet is placed between the layers to optically-isolate the horizontal fibers. 
Such a technique can be easily scaled up to layers of at least $50 \times 100~\text{cm}^2$. 
In fact, the limit in the size of the layer of glued cubes is given by the dimension of the original scintillator volume and the capacity of the CNC machine.

\begin{figure}
\centering
\raisebox{-0.5\height}{\includegraphics[width=7cm]{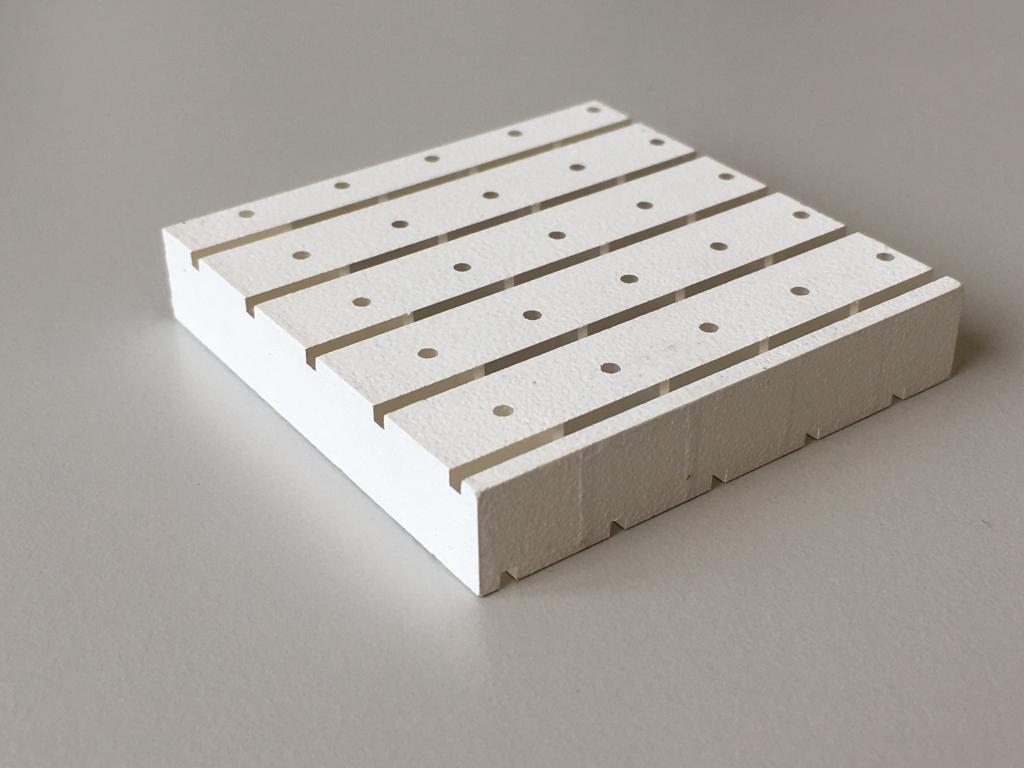}}
\raisebox{-0.5\height}{\includegraphics[width=7cm]{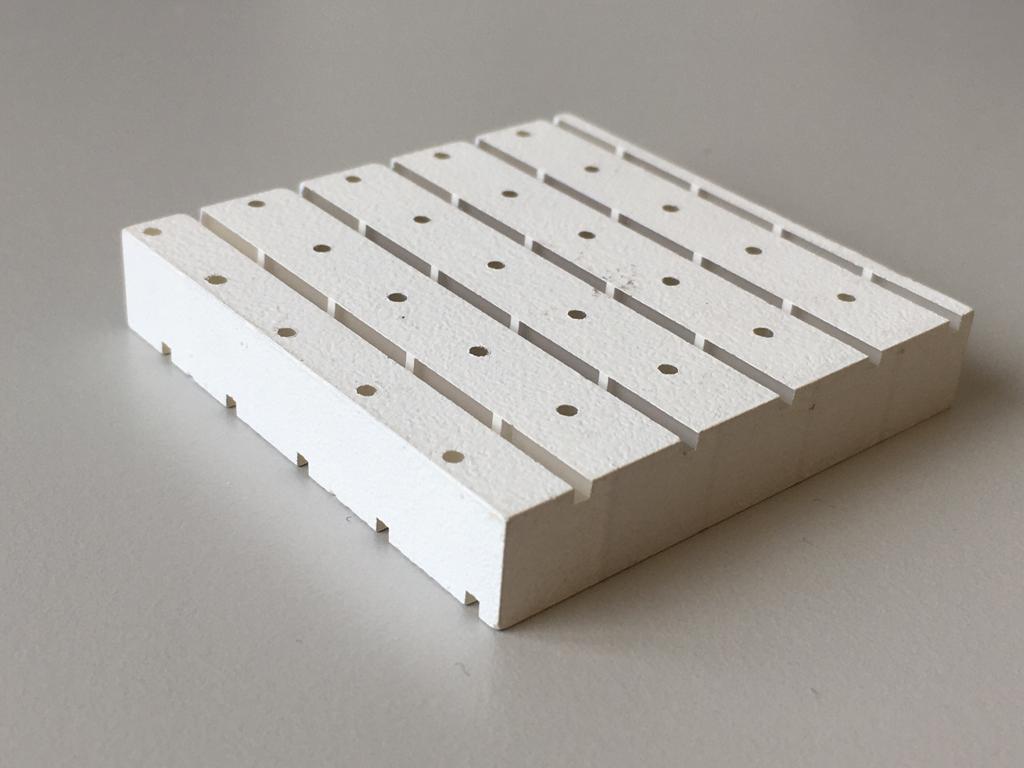}}
\raisebox{-0.5\height}{\includegraphics[width=7cm]{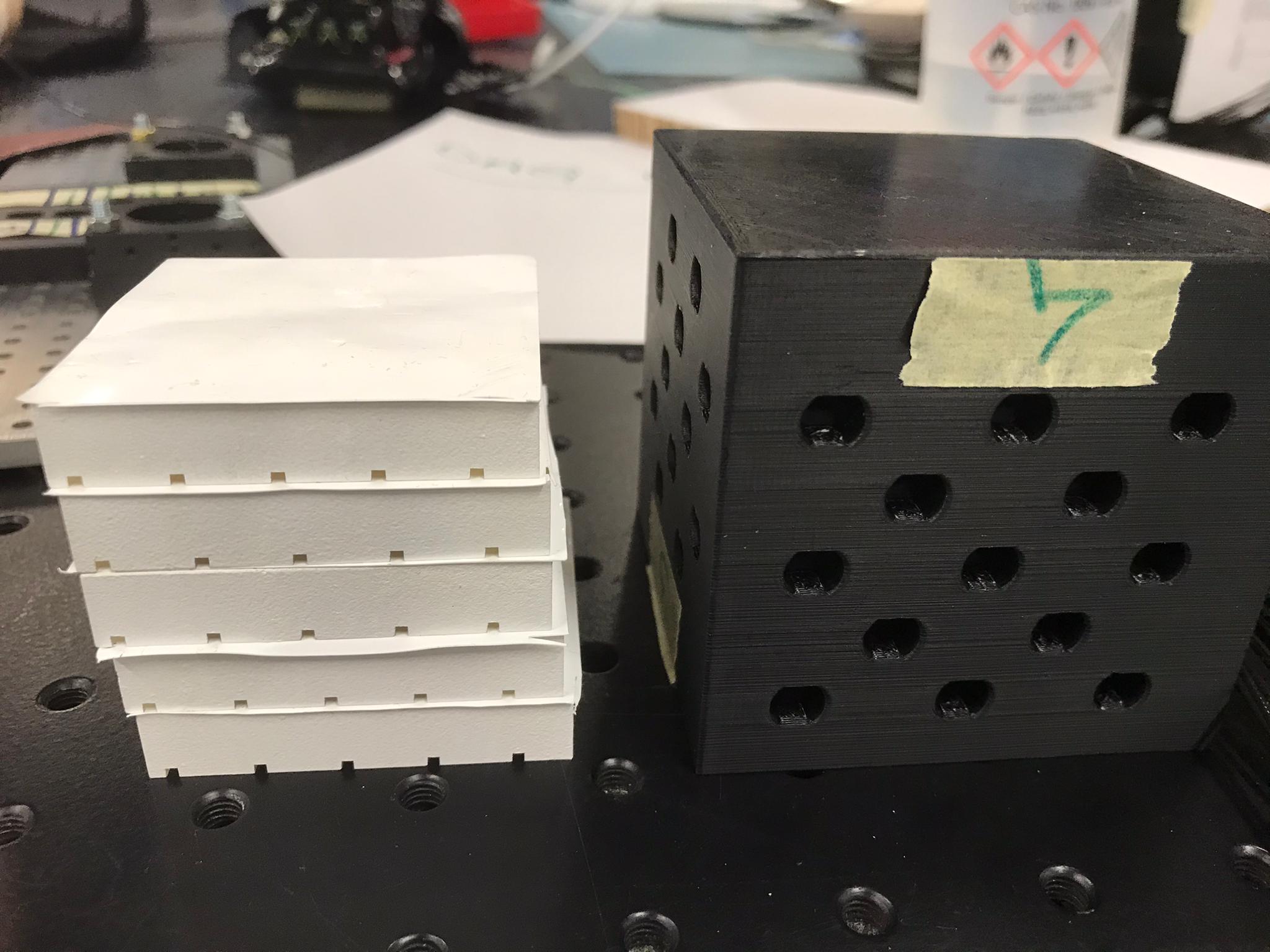}}
\caption{\label{fig:design} 
Top: Photograph of a $5 \times 5$ glued-cube layer from top (left) and bottom (right) views.
Bottom: Stack of five glued-cube layers interleaved with thin white tyvek sheets. 
The groove is 1.6 (width) $\times$ 1.5 (depth) $\text{mm}^2$ and its center is positioned at 3~mm from one edge.
}
\end{figure}

In the following section the detector performance of this ``glued-cube'' design is evaluated using a prototype made from a stack of five layers of $5 \times 5$ glued cubes (also shown in Fig.~\ref{fig:design}).
Three layers were instrumented, each one with a total of six readout channels.

%
%

%
%

\section{Performance of glued cubes}
\label{sec:measurements}

The performance requirements for a plastic scintillator detector include: high scintillation light yield, high transparency, fine granularity (if combined with optical isolation), long-term stability and fast scintillation light production. 
Although the prototype design described in Sec.~\ref{sec:design} enables the easy assembly of many segmented scintillator objects in a 3D matrix, the plastic scintillator is made of exactly the same composition and quality as the one produced by ISMA, which is used for commercial and high-energy physics applications~\cite{Artikov:2005mg,Baranov:2018uxp}.
It is therefore reasonable to expect that the scintillation performance of the prototype described will fulfill the requirements for use in future experiments.
Some differences may be introduced by the particular configuration of the grooves and quality of the optical reflector, which could undergo a future optimization based on particular experimental requirements.

In this section we aim to verify the performance of the prototype matrix by evaluating both the scintillation light output and the probability of the scintillation light to migrate from one cubic volume to an adjacent one (optical crosstalk) using cosmic-ray data. 
Whilst a low light output would affect the particle energy loss resolution, by introducing large fluctuations in measured energy deposits, a relatively large crosstalk
would reduce the detector tracking capability and complicate the separation of neighbouring particle tracks.
This is particularly true if highly-ionizing particles, such as low-energy hadrons, are involved.

Kuraray Y11 single-cladding WLS fibers~\cite{kuraray-catalogue} with a 1~mm diameter were used to capture the scintillation light, shift it to the green band and convey it to 
Hamamatsu S13360-1350CS Multi-Pixel Photon Counters (MPPC). 
The WLS fibers are approximately 5~cm long so the light attenuation is negligible.
The number of photoelectrons (PE) was extracted after measuring the MPPC gain, defined as the number of ADC counts to the number of PE conversion. The MPPC active region ($1.3\times1.3~\text{mm}^2$) was directly coupled to the polished WLS fiber end using black optical connectors, following a design similar to Ref.~\cite{Antonova:2017cdw}. 
A piece of soft black EPDM foam was placed inside the connector to push the MPPC against the WLS fiber end and improve the coupling.

\begin{figure}
\centering
\raisebox{-0.5\height}{\includegraphics[width=9cm]{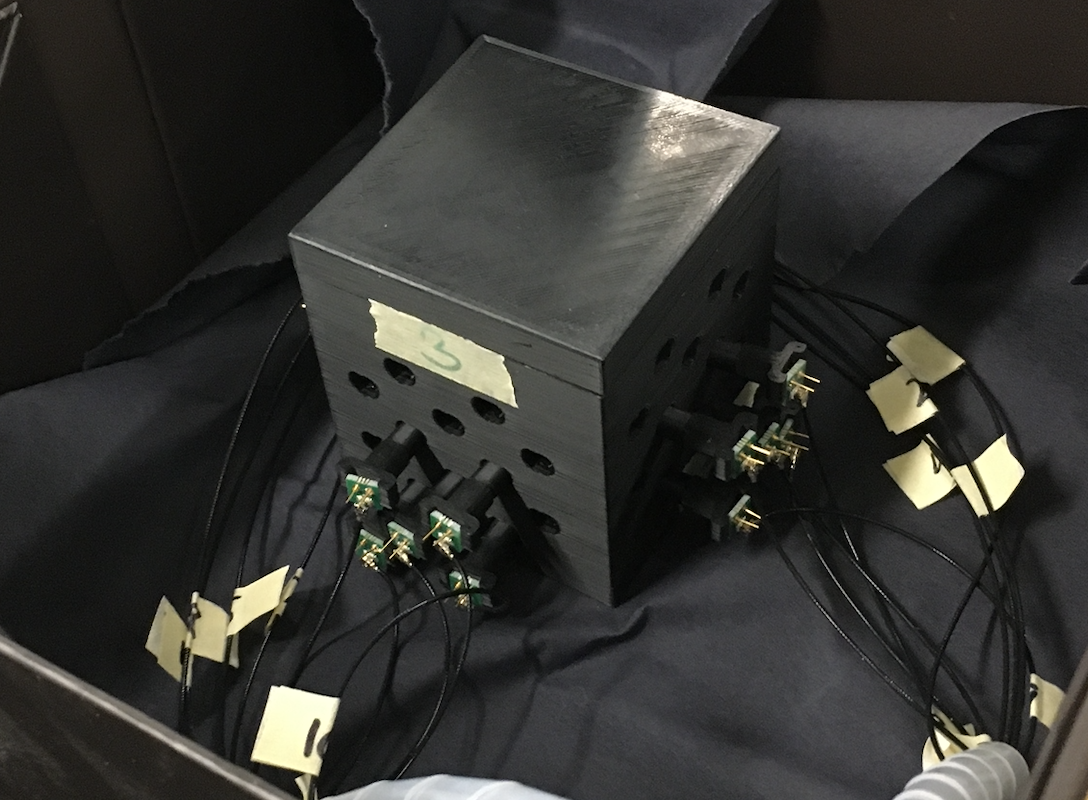}}
\caption{\label{fig:setup} Setup used to evaluate the performances of glued-cube layers.}
\end{figure}

After inserting the 
scintillator
layers into a light-tight 3D printed dark box, the WLS fibers are inserted into the grooves of the glued-cube layers and the optical connectors fixed into holes on the dark box walls (with a single-ended readout).
The whole apparatus is then placed inside another dark box and covered with black cloth.
In order to fit the high density of channels for each readout view (separated by 1~cm), both sides of the dark box were instrumented in a staggered configuration to allow a 2~cm distance between adjacent optical connectors.
The resultant setup is shown in Fig.~\ref{fig:setup}. 
The MPPC signal is read-out with a CAEN DT5702 front-end board \cite{caen-catalogue}.
In total the stack of three intermediate layers were instrumented for the tests.
Six instrumented channels (two views) in each of the three layers were sufficient for triggering on cosmic-ray muons and providing good quality data to evaluate both the light output and the cross talk probability.

\subsection{Light yield measurement}
\label{sec:measurements-light-yield}

\begin{figure}
\centering
{\includegraphics[width=0.48\linewidth]{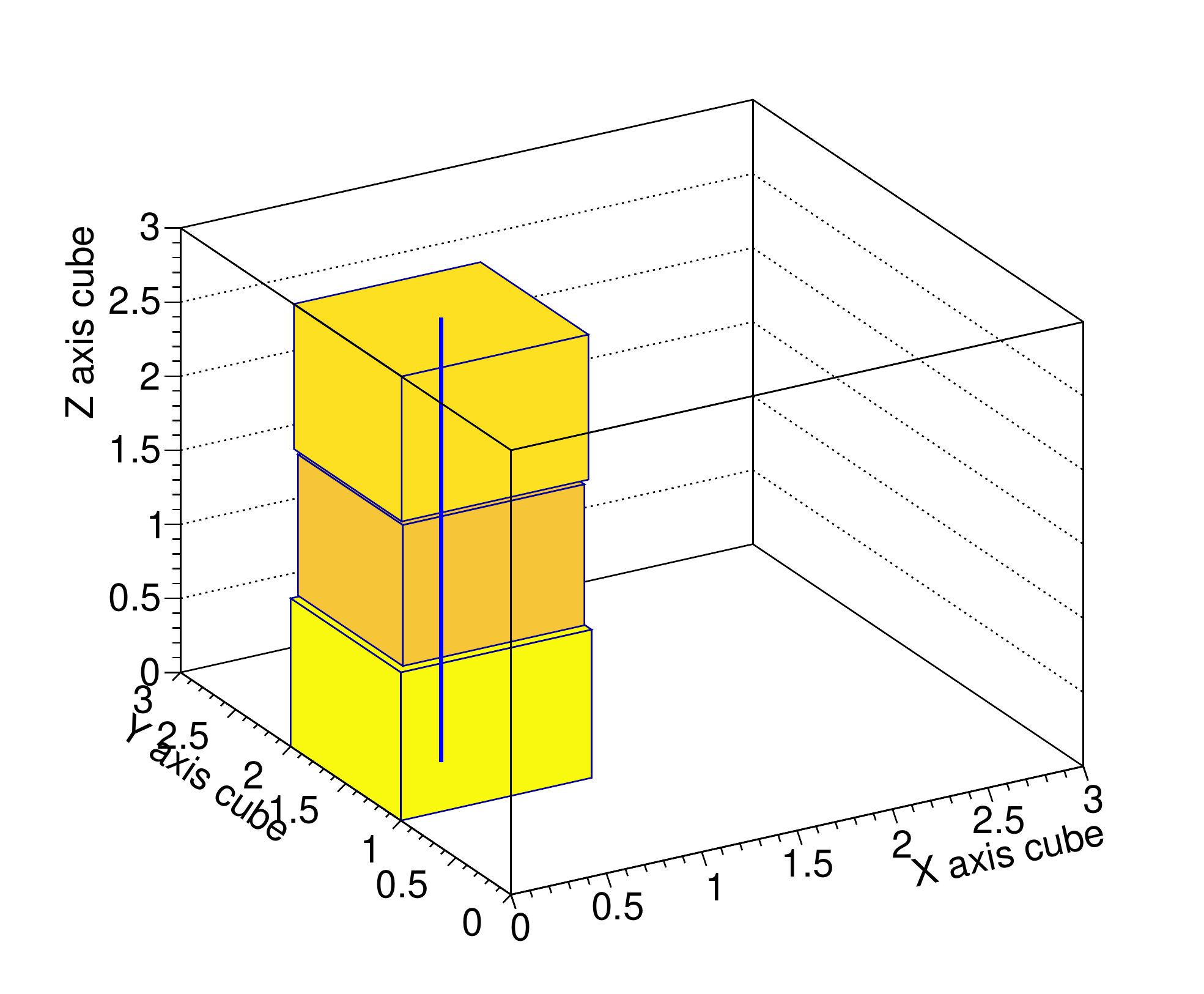}}
{\includegraphics[width=0.48\linewidth]{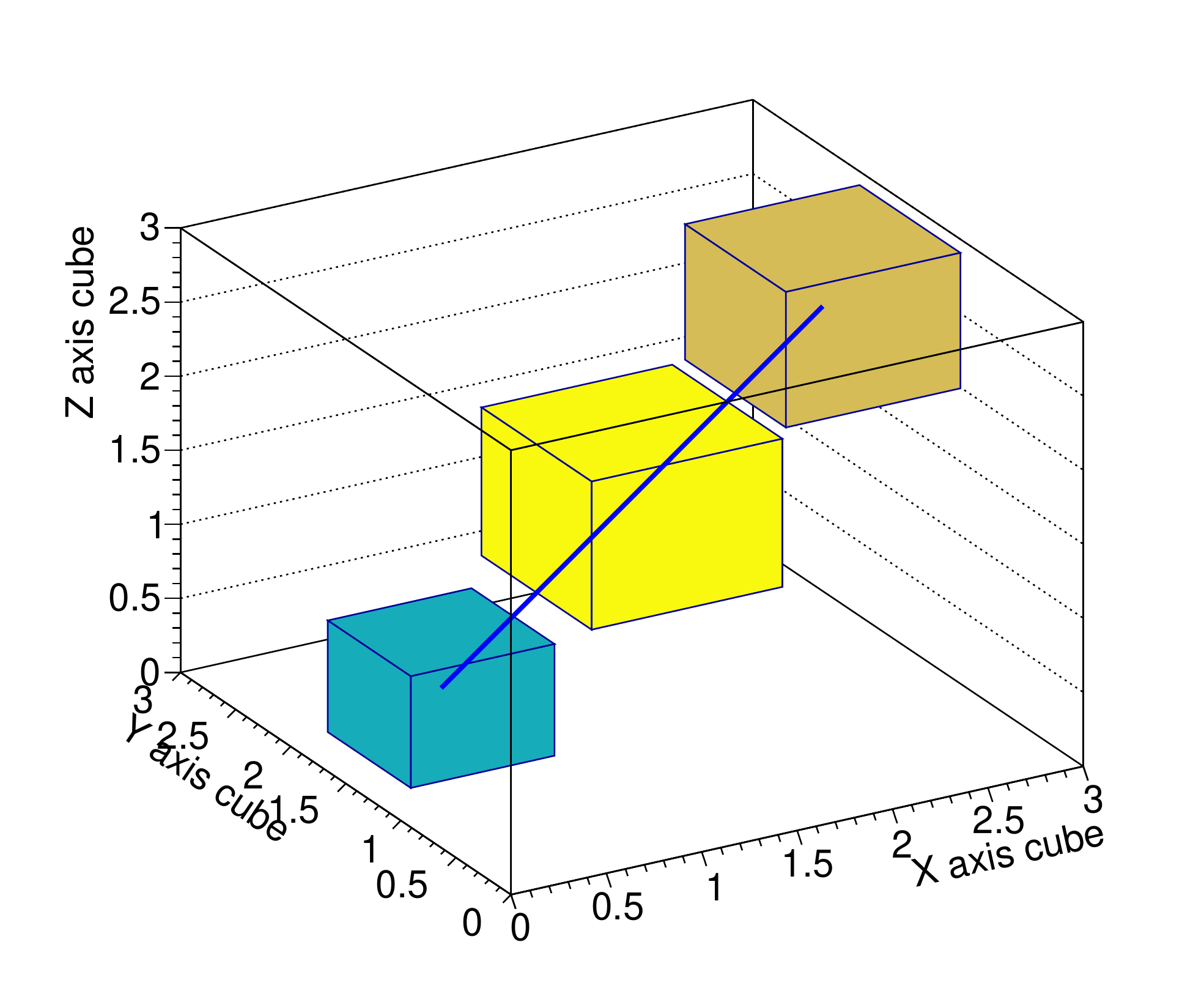}}
\caption{\label{fig:prettyCosmic} Two example event displays showing cosmic-ray muon candidates. The Z axis is orientated perpendicular to the Earth's surface. The size of boxes indicate the ADC recorded by the sum of X and Y axis channels in each layer. The blue solid line shows the fitted track trajectory. The left plot is an example of an event used in this analysis (a vertical cosmic-ray candidate).}
\end{figure}

The light yield of the scintillator is measured using cosmic ray muons identified as entering orthogonal to the top plane of the matrix. 
These muons are selected by requiring two hits on each layer, one per view, above a pre-defined ADC threshold to identify the XYZ coordinate of the cosmic crossing point.
The ADC threshold is chosen to minimise coincidences from background noise while rejecting almost no cosmic ray muons that travel the expected 1 cm through a layer. 
An example of a selected cosmic-ray muon candidate is shown in Fig.~\ref{fig:prettyCosmic}. 
Vertical muons, i.e. those crossing the three layers on the same horizontal XY position, were used for the data analysis.
A cosmic-ray induced hit gives around 2000 ADC whereas the threshold is set to 500 ADC. Following the selection, 2039 good quality cosmic ray muon candidates are considered in the analysis. 

Before the light yield can be determined, a pedestal subtraction and gain correction is made for each channel. The pedestal is established by taking the mean ADC count from events in which there is no hit above the aforementioned threshold within the channel of interest or in any adjacent channel (to avoid cross talk contributing to the estimation of the pedestal). The typical pedestal is found to be around 100-120 ADC with an RMS width of 15-30 ADC. 

The gain for each channel is defined as the number of ADC counts signifying a detected PE.
It is determined using dedicated calibration data in which the matrix is exposed to positrons from a $\text{Sr}^{90}$ source.
The scintillation light from the electrons provides high intensity data with a wide range of PE deposits. Analysing ADC counts for each channel clearly reveals discretised PE peaks, the mean distance of the peaks gives the gain. 
Typical gain values of 35-40 ADC per PE are obtained. 

\begin{figure}
\centering
{\includegraphics[width=0.49\linewidth]{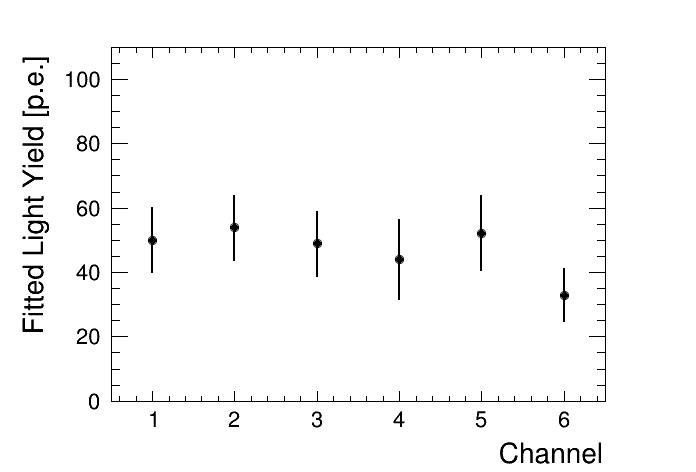}}
{\includegraphics[width=0.49\linewidth]{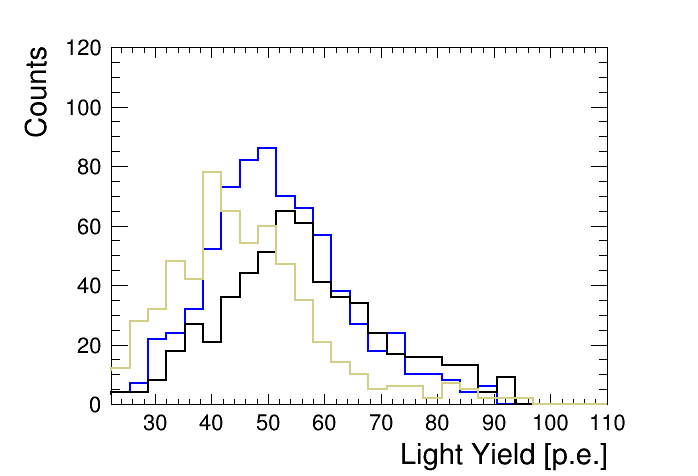}}
\caption{\label{fig:lightYield}\textbf{Left:} the results of the fit to the light yield distributions for the six central layer channels. The central value shows the mean from the fit while the error bar shows the standard deviation. \textbf{Right:} three example distributions which are fit to form the left plot. 
}
\end{figure}

Once the gain and pedestal is found for each channel, the light yield is determined by taking each identified cosmic-induced hit, subtracting the pedestal from the observed ADC and then applying the gain correction. Although this can be done for all channels, it is possible for channels in the top and bottom layers to have a cosmic ray muon enter or exit part way through a detection element (such that the path length travelled in a layer is less than 1 cm), for this reason only the six channels in the central layer are used to determine the light yield. 
The distribution of light yield is measured for each channel and a Gaussian fit is made to determine the most probable value and spread. 
The results of these fits, alongside examples of the light yield distributions, are shown in Fig.~\ref{fig:lightYield}. The dip in the last channel is likely from an imperfect coupling of the fibre to the MPPC.

\subsection{Optical crosstalk}
\label{sec:measurements-cross-talk}

The optical crosstalk within the scintillator block is measured with vertical going cosmic ray muons following the same procedures as described in section \ref{sec:measurements-light-yield}, also using only the central layer. 

\begin{figure}
\centering
{
\includegraphics[width=0.80\linewidth]{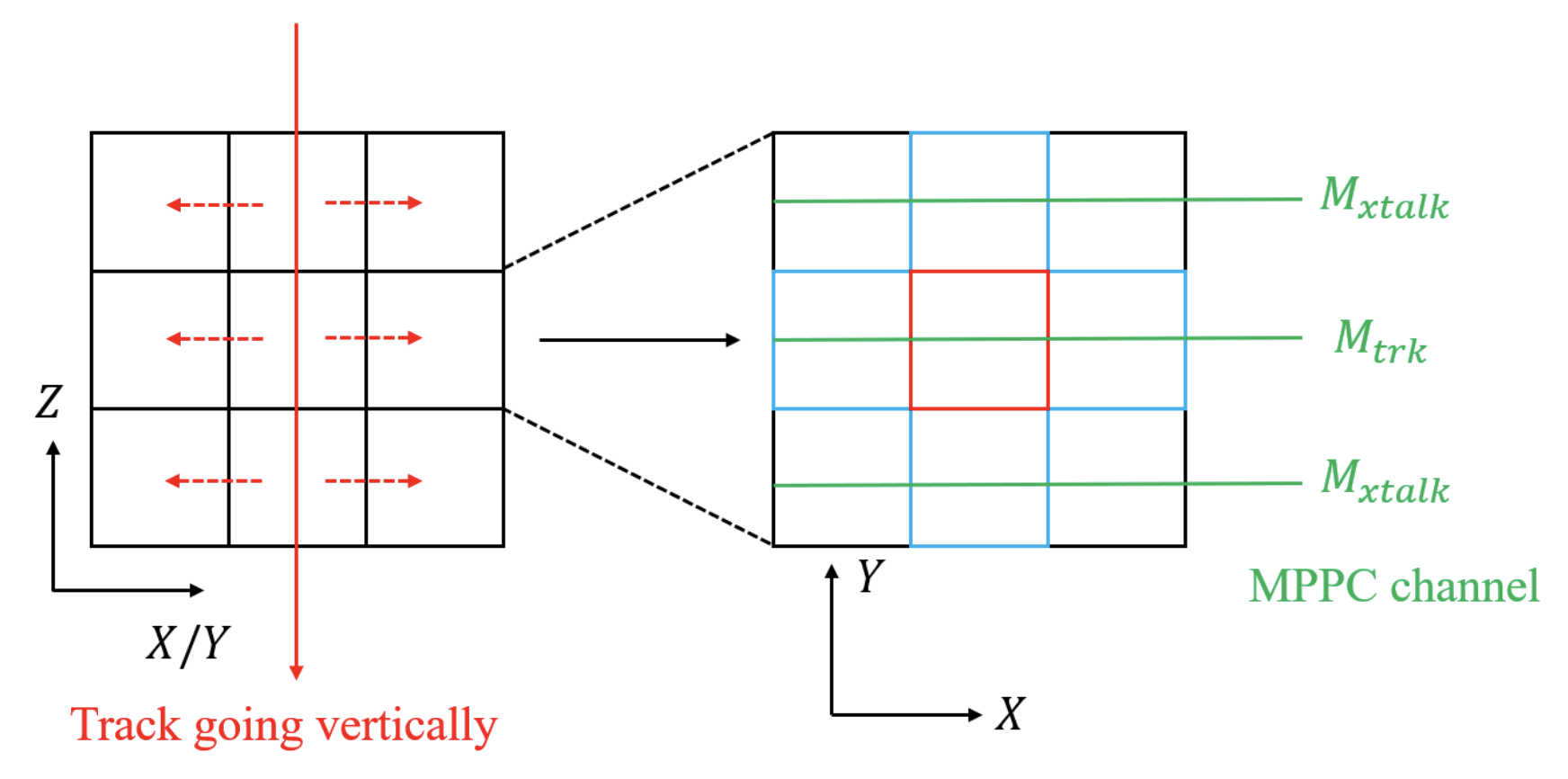}
}
\caption{\label{fig:xtalk_sketch} A sketch illustrating optical crosstalk measurement from a vertical cosmic muon passing through the matrix where the cubes show optically isolated elements. The crosstalk is measured along X-axis or Y-axis. The cubes with red boarders shows where track passes through and shows the adjacent blue cubes which are expected to be affected by crosstalk. The green lines show which detection elements correspond to which channels and which signal is used to measure the light from the primary track ($M_{trk}$) and which is expected to record the crosstalk ($M_{xtalk}$).}
\end{figure}

\begin{figure}
\centering
{\includegraphics[width=0.50\linewidth]{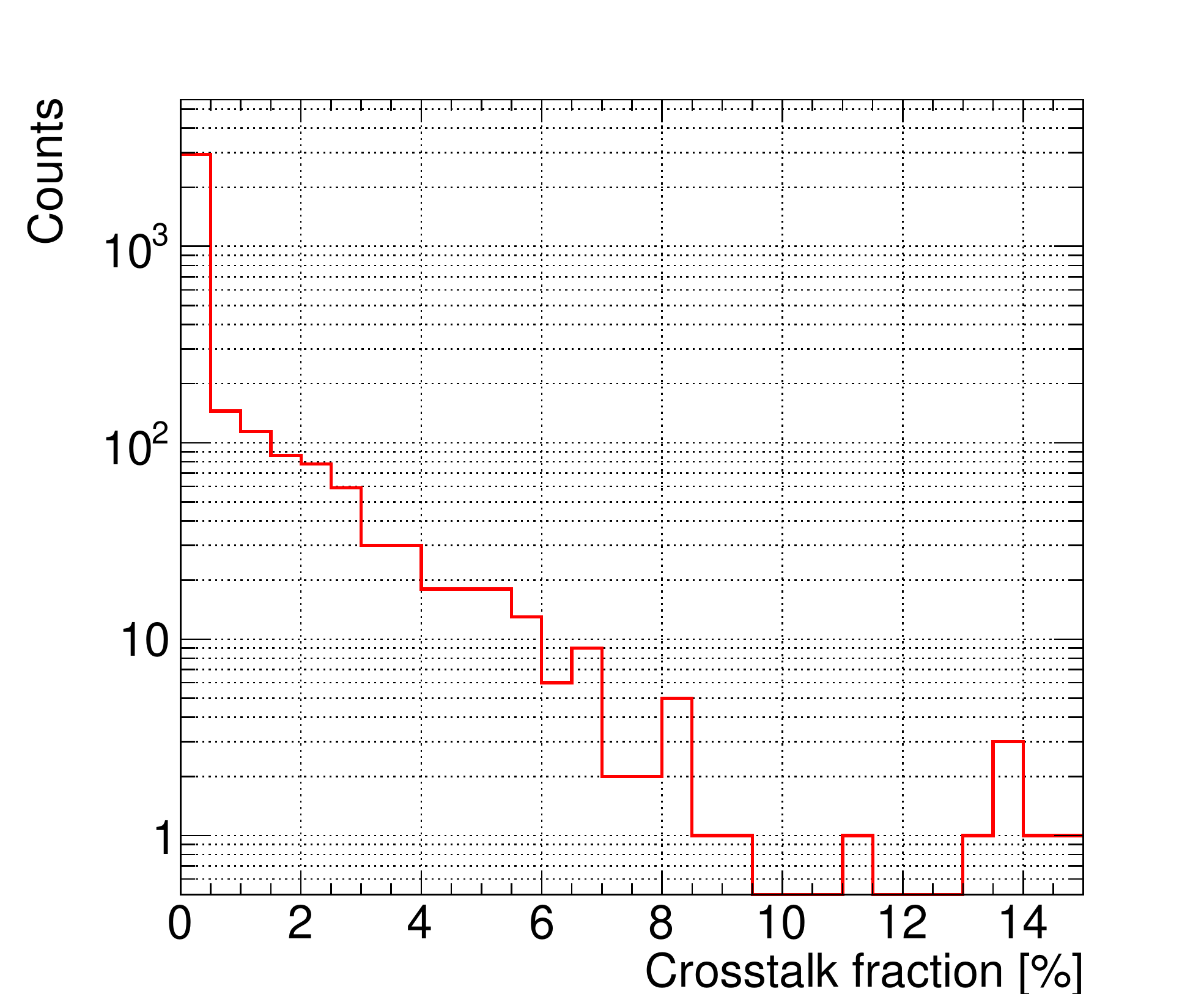}}
\caption{\label{fig:xtalk_result} Result of the crosstalk measurement. The histogram is composed of crosstalk fraction measurements made in the central layer for a large ensemble of identified cosmic ray tracks. The mean of the histogram is 0.45\%.}
\end{figure}

Fig.~\ref{fig:xtalk_sketch} details schematically how the crosstalk is measured. First the scintillation signal (pedestal subtracted and gain corrected) for the observed track is measured in each layer (denoted by $M_{trk}$). Crosstalk is then expected to occur in adjacent elements, the signal from which is measured with their corresponding channels ($M_{xtalk}$). The crosstalk fraction is measured as $M_{xtalk}/M_{trk}$ for each observation of $M_{xtalk}$ in a layer. Measurements of the crosstalk fraction are repeated for a large number of cosmic tracks and the distribution of measurements on the channels in the central layer are reported in Fig.~\ref{fig:xtalk_result}. On every channel it is found that the crosstalk was so low that it could not be distinguished from the pedestal. After accounting for systematic effects and noise in the setup, the crosstalk is unlikely to be greater than 2\%.

\subsection{Validations}
\label{sec:validations}

To validate the experimental setup used to obtain these results, the same measurements performed in Sec.~\ref{sec:measurements} were repeated using an analogous matrix containing 1~cm$^{3}$ cubes, of the same type as discussed in Ref.~\cite{superfgd}. The resultant light yield was found to be in the range of 70-80 PE and the cross talk at the level of $\sim$3\% (in this case it was easily distinguished from the pedestal). These results are broadly also compatible with those reported in Ref.~\cite{superfgd}, but it should be noted that the experimental setups are not identical. 



%
%

\section{Discussions}
\label{sec:discussions}

The results indicate a light yield of $\sim$50 PE for a minimum ionising particle traversing 1 cm. 
This is comparable to the $\sim$70 PE obtained using 1~cm$^{3}$ scintillating cubes 
produced by Uniplast in Vladimir (Russia)
for use in the future T2K's upgraded near detector~\cite{superfgd}.
One possible explanation for difference could be the different doping content in the polystyrene that, if needed can be further optimised to increase the light yield.
In any case, the light yield per readout channel remains higher or analogous to that of scintillator detectors currently used in neutrino experiments. Some examples can be found in Ref.~\cite{t2k-fgd,SoLid:2018lkh,DANSS:2018fnn}. 
Updating the glued-cube layer groove design from square to circular may allow an increase in both the absolute light yield and the light yield uniformity among channels in each layer, thanks to the improved contact that reduces the air gap between the plastic scintillator and the WLS fiber. 
It is also foreseeable that the lateral position of the grooves with respect to the center of the cube can be further optimised.

Further improvements may also be made by reducing the groove size from 1.5~mm to$\sim$1~mm to better match the diameter of the WLS fiber. The total light output can be additionally enhanced by uniformly placing optical glue in the grooves to remove the air gap between the scintillator and the fiber, responsible for the reflection of a relatively large amount of photons (up to 50\% in certain conditions).
Although these updates are only applicable to the grooves in the glued-cube layers and not the holes, they are expected to improve the overall light output performance without creating any particular issues for the detector assembly.

The crosstalk is estimated to be less than 2\%, which is comparable to the state-of-the-art reported in 
~\cite{superfgd}.
This is expected from the thicker reflector layers that are used between neighbouring elements in each glued-cube layer.
Both the epoxy glue and the $\text{TiO}_2$ paint are white reflector widely used in scintillator detectors.
If necessary, a thinner glue reflector could be used without expecting a dramatic increase in the crosstalk.

Though there is potential to further improve the performance of the prototype tested here, the results already demonstrate that alternative scintillator manufacturing methods can be used to provide comparable performance to using 1~cm$^{3}$ scintillating cubes, whilst vastly reducing the complexity of detector assembly. 
Although results with a small scintillator sample has been reported, such a technique can be easily scaled up to at least $50 \times 100~\text{cm}^2$, i.e. a single block of 5,000 optically-isolated $1~\text{cm}^3$ cubes. It is possible to envisage the production of a rigid 3D matrix from the individual layers. For example, a thin layer of glue or bi-adhesive reflecting tyvek could be placed between layers during assembly.

Overall, the method proposed in this manuscript enhances the feasibility of producing and assembly compact scintillator detectors made of a very-large number of optically-isolated 3D volumes. 
This in turn allows an easier scaling of fine-granularity scintillator detectors to larger masses and to more easily produce detectors with smaller 3D optically-isolated volumes.

%
%

\section{Conclusions}
\label{sec:conclusions}

In this article we have reported the results of a feasibility study, demonstrating the production of large single-blocks of plastic scintillator made of 
many
tiny optically-isolated cubes glued together and assessing their potential performance in a particle detector.
Overall, it is shown that the scintillator performances are comparable to the state of the art while providing the means for an easier production and assembly of future 3D fine-granularity scintillator detectors with larger masses.
Although this work reports on the performance of a relatively small prototype, the proposed production method can be easily scaled up to glued-cube layers of at least $50 \times 100~\text{cm}^2$.
It is important to note that existing technology, such as the 1~cm$^{3}$ cubes design, \textit{can} be scaled to very large detectors but that the assembly and production would be much slower, more technically challenging, and likely also more costly compared to simply stacking the single-block layers exhibited in this article.  
Hence, this work may be particularly relevant for the future scintillator-based neutrino detectors that will require both a fine 3D granularity and a target mass larger than that used in currently running experiments.



\bibliographystyle{utphys}
\bibliography{GluedCubes.bib}

\end{document}